\documentclass[amstex,aps,preprint,nofootinbib,floatfix]{revtex4}%

\usepackage{amssymb}
\usepackage{graphicx}%

\begin{document}

\title{EPR correlations without the EPR dilemma: a local scheme}
\author{A.\ Matzkin }

\address{Laboratoire de Spectrom\'{e}trie physique (CNRS Unit\'{e} 5588),
Universit\'{e} Joseph-Fourier Grenoble-1, BP 87, 38402 Saint-Martin
d'H\`{e}res, France}
\begin{abstract}
A model for two entangled systems in an EPR setting is shown to reproduce
the quantum-mechanical outcomes and expectation values. Each system is
represented by a small sphere containing a point-like particle embedded in a
field.\ A quantum state appears as an equivalence class of several possible
particle-field configurations. Contrarily to Bell-type hidden variables
models, the fields account for the non-commutative aspects of the
measurements and deny the simultaneous reality of incompatible physical
quantities, thereby allowing to escape EPR's ``completeness or locality''
dilemma.
\end{abstract}

\maketitle

\section{Introduction}

In their celebrated paper \cite{EPR}, Einstein, Podolsky and Rosen (EPR)
argued that quantum mechanics was incomplete on the ground that for
entangled states the formalism predicts with certainty the measurement
outcomes of noncommuting observables although they cannot have simultaneous
reality. They argued the alternative to incompleteness was to make the
reality of one particle's properties depend on the measurement made on the
other particle, irrespective of their spatial separation.\ EPR concluded:
\textquotedblleft\emph{no reasonable definition of reality could permit this}%
\textquotedblright\ nonlocal action at-a-distance. In a seminal work \cite%
{bell1964}, Bell showed that local models based on a distribution of hidden
variables (HV) intended to complete quantum mechanics must satisfy an
inequality involving averages taken over the hidden variable distributions.
He also showed that in certain circumstances the average values of
two-particle quantum observables violate these inequalities. However, it is
seldom mentioned that Bell-type models are only a subset of the local models
that can be envisaged. Indeed, Bell's theorem \cite{CHSH,bell2} is grounded
on two important assumptions: (a) the HV\ ascribe a sub-quantum elementary
probability for any 1 or 2-particle outcomes; (b) this probability
factorizes into two single particle probabilities. These assumptions lead
\cite{fine82} to the existence of a \emph{joint} probability function for
all the observables entering the inequality (though there is no such
probability according to quantum mechanics), thereby accounting for the
'simultaneous reality' appearing in the EPR dilemma.\ General arguments seem
to indicate that these assumptions are needed to comply with the EPR
requirements but are by no means necessary ingredients in order to enforce
locality \cite{jaynes,orlov,rec2}. In this work we put forward a scheme
compatible with quantum mechanical correlations but does not abide by the
EPR dilemma. The model, developed for the prototypical spin-1/2 pair,
describes each system by postulating a particle and a classical field. It is
shown that different particle-field configurations yield the same
probabilities for outcomes detection, even when the outcomes can be
predicted with certainty. We will first put forward the model for a \emph{%
single} particle. We will then naturally extend the model to the
two-particle case, and show how, by introducing a correlation \emph{at the
source}, the model reproduces the quantum predictions that violate the Bell
inequalities without involving action at a distance.

\section{Single field-particle system}

Let a single spin-1/2 be represented by a field-particle system assumed to
be composed of a small sphere, with the position of its center in the
laboratory frame being denoted by $\mathbf{x}$ and the internal spherical
variables relative to the center of the sphere by $\mathbf{r}\equiv
(r,\theta ,\phi )$ (see Fig. 1a). A classical scalar field $F(\mathbf{r})$
is defined on the spherical surface.\ The point-like particle sits still at
a fixed (but unknown position) on the sphere. We are interested in measuring
the polarization of the particle, i.e., its internal angular momentum
projection along a given axis\footnote{%
From a physical standpoint, what we have called here the position of the
particle should more properly be called the position of the particle's
angular momentum $\mathbf{r}_{0}\times \mathbf{p}$ relative to the center of
the sphere. We will not make explicitly this distinction in this paper.}.
Let $\varepsilon _{b}$ denote the polarization along an axis $b$ making an
angle $\theta _{b}$ with the $z$ axis. We assume that the possible outcomes $%
\varepsilon _{b}=\pm 1$ can be obtained, the result depending in a manner to
be described below (i) on the region on which the field is defined
and (ii) on the position of the particle. The elementary support of the field
$F$ is a hemispherical surface. The value of
the field at any point depends on the projection of that point on the
axis. Let $\Sigma _{+a}$ denote the positive half-sphere centered on the
axis $a$ making an angle $\theta _{a}$ with the $z$ axis, and $F_{\Sigma
_{+a}}$ denote the field distributed on that hemisphere. $F_{\Sigma _{+a}}(%
\mathbf{r})$ is defined by%
\begin{equation}
F_{\Sigma _{+a}}(\mathbf{r})=\left\{
\begin{tabular}{l}
$\mathbf{r}\cdot \mathbf{a}e^{i\phi _{+a}}/\pi R^{2}$ if $\mathbf{r}\in
\Sigma _{+a}$ \\
$0$ otherwise%
\end{tabular}%
\ \ \ \ \ \right. ,  \label{2}
\end{equation}%
$R$ being the radius of the sphere and $\phi $ the phase of the field; for
simplicity we will take all the axes to be coplanar with $z$ and the phase
will be assumed constant over an entire hemisphere (the phase thus appears
as a global additional degree of freedom of the field). The mean value of $%
\mathbf{r}\cdot \mathbf{b}/\pi R^{2}$ taken over $\Sigma _{+a}$ is given by
\begin{equation}
\left\langle F_{\Sigma _{+b}}+F_{\Sigma _{-b}}\right\rangle _{\Sigma
_{+a}}\equiv \int_{\Sigma _{+a}}\frac{\mathbf{r}\cdot \mathbf{b}}{\pi R^{2}}d%
\mathbf{\hat{r}}=\cos \left( \theta _{b}-\theta _{a}\right) ,  \label{3}
\end{equation}%
where $d\mathbf{\hat{r}}$ denotes the spherical surface element for a sphere
of radius $R$ and we have set $\phi _{\pm b}=0$. The only requirement we
make on the particle's position is that it must be embedded within the field:
the particle cannot be in a field free region of the sphere.

When the polarization $\varepsilon _{b}$ is measured we postulate that the
measuring apparatus along $b$ interacts with the field $F_{\Sigma _{+a}}.$
Let $[a+b]$ and $[a-b]$ denote the directions lying halfway between the axes
$a$ (of the distribution) and $b$ or $-b$ (of the measuring direction), with
respective angles $(\theta _{b}+\theta _{a})/2$ and $(\theta _{b}+\pi
+\theta _{a})/2$. We will assume that the field-apparatus interaction
results in a \emph{rotation} of the original pre-measurement field $%
F_{\Sigma _{+a}}$ toward both of the apparatus axes, $F_{\Sigma
_{+a}}\rightarrow \left( F_{\Sigma _{+b}}+F_{\Sigma _{-b}}\right) e^{i\phi
_{+a}}$ (Fig 1b); $\phi _{+a}$ is the phase of the original field and we
will suppose the measurement does not introduce additional phases. A
definite outcome $\varepsilon _{b}=\pm 1$ depends on which of the
hemispheres $\Sigma _{\pm b}$ the particle is after the interaction. In
terms of the field, this probability is given by the relative value of the
average of the rotated field $F_{\Sigma _{+b}}+F_{\Sigma _{-b}}$ over the
intermediate 'half-rotated' hemisphere $F_{\Sigma _{\lbrack a\pm b]}}$, yielding in accordance
with Eq. (\ref{3})%
\begin{eqnarray}
P_{F_{\Sigma _{+a}}}(\varepsilon _{b}& =+1)=\left\vert \left\langle
F_{\Sigma _{+b}}+F_{\Sigma _{-b}}\right\rangle _{\Sigma _{\lbrack
a+b]}}\right\vert ^{2}/N=\cos ^{2}\frac{\theta _{b}-\theta _{a}}{2}
\label{10} \\
P_{F_{\Sigma _{+a}}}(\varepsilon _{b}& =-1)=\left\vert \left\langle
F_{\Sigma _{+b}}+F_{\Sigma _{-b}}\right\rangle _{\Sigma _{\lbrack
a-b]}}\right\vert ^{2}/N=\sin ^{2}\frac{\theta _{a}-\theta _{b}}{2}
\label{12}
\end{eqnarray}%
with $N\ $being the sum of both terms, thereby recovering the probabilities
of measurements made on a single spin-1/2, reading in the standard notation $%
\left\vert \left\langle \pm b\right. \left\vert +a\right\rangle \right\vert
^{2}$ (normalization will be implicitly understood in the rest of the
paper). If $b$ and $a$ are taken to be the same, then one has $\Sigma
_{\lbrack a+a]}\equiv \Sigma _{+a}$ and $P_{\Sigma _{+a}}(\varepsilon
_{a}=\pm 1)=1$ and $0$ respectively. Hence a field $F_{\Sigma _{+a}}$
corresponds to a well-defined positive polarization along the $a$ axis. In
this case the symmetry axis of the field distribution coincides with the
measurement axis and the system-apparatus fields interaction has no effect:
the particle's pre and post-measurement position remains within the same
hemisphere $\Sigma _{+a}$. The particle's hemispherical position can be said
to determine the outcome, i.e. %\begin{equation}
$P_{F_{\Sigma _{+a}}}(\varepsilon _{a}=\pm 1)=P_{F_{\Sigma _{+a}}}(\mathbf{%
r\in }\Sigma _{\pm a})=1$ or $0$. %\frac{1}{2}\pm\frac{1}{2}.   \label{16}
%\end{equation}
On the other hand when $b$ and $a$ lie along different directions, the
particle position cannot ascribe probabilities: the probabilities depend on
the system and apparatus fields and $\varepsilon _{b}$ only acquires a value
$\pm 1$ \emph{after} the system field has interacted with the apparatus and
rotated toward the measurement axis. A straightforward consequence is that
the measurements do not commute, and thus joint polarization measurements
along different axes are undefined.

Since fields obey the principle of superposition, we can envisage
superpositions of fields defined on different hemispheres. But fields
defined on different hemispheres turn out to be \emph{equivalent }to a field
defined on a single hemisphere. Indeed it is easy\footnote{%
As the reader will have noted, the fields are defined through a mapping of
the Hilbert space rays onto the relevant hemispherical surface.} to see that
one can write for any axis $u$
\begin{equation}
F_{\Sigma _{+a}}\sim \cos (\frac{\theta _{u}-\theta _{a}}{2})F_{\Sigma
_{+u}}+\sin (\frac{\theta _{u}-\theta _{a}}{2})F_{\Sigma _{-u}},  \label{z1}
\end{equation}%
meaning that although the two fields on the right and left handsides (hs) of
Eq. (\ref{z1}) are different -- they are not defined on the same
hemispherical surfaces --, they lead to exactly the same predictions.
Indeed, when measurements are made along \emph{any} axis $b$ the averages of
the left and right hs of Eq. (\ref{z1}) give the same result $\cos (\frac{%
\theta _{a}-\theta _{b}}{2})$. These fields thus define an \emph{equivalence
class.} From the particle standpoint a definite field configuration implies
a different behaviour: for the\ field on the rhs of Eq. (\ref{z1}), denoted $%
F_{rhs}$, the no-perturbation axis is $u$, not $a$, and the particle
distribution cannot be uniform. Hence there is a probability function $%
p_{F_{rhs}}(\varepsilon _{u}=\pm 1,\mathbf{r})=1$ or $0$ depending on
whether $\mathbf{r\in }\Sigma _{\pm u}$ and such that $P_{F_{rhs}}(%
\varepsilon _{u}=\pm 1)$ is recovered by integration over the particle
distribution. %\begin{equation}
%P_{F_{rhs}}(\varepsilon _{u}=\pm 1)=\int p_{F_{rhs}}(\varepsilon _{u}=\pm 1,%
%\mathbf{r})\rho _{rhs}(\mathbf{r})d\mathbf{r}=\cos ^{2}\left( \frac{\theta
%_{u}-\theta _{a}}{2}+\frac{\pi }{4}(1\pm 1)\right) ,  \label{z2}
%\end{equation}%
%where $\rho _{rhs}(\mathbf{r})$ denotes the particle distribution when the
%field is given by the rhs of Eq. (\ref{z1}).
For $b\neq u$ however there is no probability function $p_{F_{rhs}}(%
\varepsilon _{b}=\pm 1,\mathbf{r})$ hence $P_{F_{rhs}}(\varepsilon _{b}=\pm
1)$ cannot depend on $\mathbf{r}$: there is no sub-field mechanism that
determines the outcome.\ This is consistent with Eqs. (\ref{10})-(\ref{12})
in which the field rotation does not allow to define joint probabilities of
the type $P_{F_{rhs}}(\varepsilon _{u}=\pm 1\cap \varepsilon _{b}=\pm 1)$;
it can be shown instead that such joint probabilities would follow by
allowing the particle position to determine probabilities for measurements
along arbitrary axes \cite{matzkin jpa08}. In the specific case of measuring
$\varepsilon _{a}$ in the field $F_{rhs}$, the system and apparatus fields
must interfere in such a way as to obtain $%\begin{equation}
P_{F_{rhs}}(\varepsilon _{a}=-1)=0%\label{z3}
%\end{equation}
$, irrespective of the initial particle's position. Finally let us introduce
the fields $F_{\alpha (u)\pm }$ defined by
\begin{equation}
F_{\alpha (u)\pm }(\mathbf{r})=e^{i\pm \frac{\pi }{2}}F_{\Sigma _{+u}}(%
\mathbf{r})+F_{\Sigma _{-u}}(\mathbf{r}),  \label{57}
\end{equation}%
which obey the equivalence $F_{\alpha (u)\pm }\sim F_{\alpha (b)\pm }$ for
any axes $u$ and $b$.\ We have
\begin{equation}
P_{\alpha (u)\pm }(\varepsilon _{b}=\pm 1)=\left\vert \left\langle F_{\alpha
(u)\pm }\right\rangle \right\vert ^{2}=\frac{1}{2}  \label{41}
\end{equation}%
for \emph{any} $b$, the average depicting Eq. (\ref{3}) taken on the rotated
hemispheres $\Sigma _{\lbrack u\pm b]}$ (for $F_{\Sigma _{+u}}$) and $\Sigma
_{\lbrack -u\pm b]}$ (for $F_{\Sigma _{-u}}$). An interpretation in terms of
the particle position can only be given for $b=u$ with elementary
probabilities $p_{F_{\alpha (u)\pm }}(\varepsilon _{u}=\pm 1,\mathbf{r})=1$
or $0$ depending on whether $\mathbf{r\in }\Sigma _{\pm u}$.\ It is
nevertheless possible to postulate additional sub-quantum probabilities
provided they are consistent with the field averages.\ For example we will
suppose for either of the fields $F_{\alpha (u)\pm }$ that
\begin{equation}
P_{\alpha (u)}(\varepsilon _{b}=1|\mathbf{r}\in \Sigma _{\pm u})=\cos
^{2}\left( \frac{\theta _{u}-\theta _{b}}{2}+\frac{\pi }{2}(1\mp 1)\right)
\label{42}
\end{equation}%
which assuming $\mathbf{r}$ is uniformly distributed is consistent with (\ref%
{41}) given that
\begin{equation}
\sum_{\pm }P_{\alpha (u)}(\varepsilon _{b}=1|\mathbf{r}%
\in \Sigma _{\pm u})P_{\alpha (u)}(\mathbf{r}\in \Sigma _{\pm u})=P_{\alpha
(u)}(\varepsilon _{b}=1).
\end{equation}
Note that Eq. (\ref{42}) supplements Eq. (\ref{41}%
) with a condition on the hemispherical position of the particle, but the
latter does not determine the outcome (it is not an elementary probability).

\begin{figure}[tb]
\includegraphics[height=1.1in,width=3.5in]{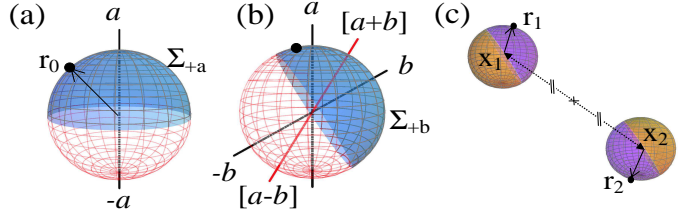}
\caption{ (a) A system is represented by a point-like particle lying on the
surface of a small sphere, on which a field is defined. The particle
position $\mathbf{r}_{0}$ is unknown; the field is defined on the hemispherical
surface centered on the positive $a$ axis. (b) Post-measurement situation after the
polarization of the system pictured in (a) has been measured along the $b$
axis, yielding an outcome $\protect\varepsilon _{b}=+1$: the field has
rotated and is now centered on $b$. (c) An initial 2-particle system has
fragmented into 2 subsystems (at $\mathbf{x}_{1}$ and $\mathbf{x}_{2}$),
each carrying a particle embedded in a field $F_{\protect\alpha (u)\pm }$.
As a result of the fragmentation the particle positions lie on opposite
points of their sphere and the fields become effectively correlated between
opposite hemispheres of each subsystem, as symbolized by the colouring.}
\label{f1}
\end{figure}

\section{Two-particle system}

Assume now an initial two-particle system is fragmented into two subsystems
flying apart in opposite directions. Each of the two particles, labeled $1$
and $2$, is embedded in a field defined on the surface of a small sphere. $%
\mathbf{x}_{1}$ (resp. $\mathbf{x}_{2}$) denotes the position of the
subsystem 1 (resp.\ 2) sphere in the laboratory frame.\ The internal
variables within each sphere are labeled by $\mathbf{r}_{1}$ and $\mathbf{r}%
_{2}$. As soon as the fragmentation process is completed, the positions of
each point-like particle as well as the fields are fixed, the polarization
of each system depending on the field distribution and the particle position
on its spherical surface. We will choose the initial correlation to
correspond to the compound having zero polarization at least along an axis $u
$ (but see below), in view of reproducing the statistics for the two
spin-1/2 in the singlet state problem. Assuming the total polarization is
conserved, the fields and particle positions must be initially correlated
such that $\varepsilon _{u}^{1}=-\varepsilon _{u}^{2}$. Let us start by
examining the \emph{no-perturbation} measurements. In this case the particle
positions determine the outcomes, from which it follows that we must set%
\begin{equation}
\mathbf{r}_{1}=-\mathbf{r}_{2}  \label{28}
\end{equation}%
at the source. Assume subsystem 1 and 2 fields to be given by $F_{\alpha
(u)+}$ and $F_{\alpha (u)-}$ defined above. The total field for the system
is thus
\begin{equation}
F_{T(u)}(\mathbf{r}_{1},\mathbf{r}_{2})=F_{\alpha (u)+}^{1}(\mathbf{r}%
_{1})F_{\alpha (u)-}^{2}(\mathbf{r}_{2}).  \label{31}
\end{equation}%
Single outcome probabilities $P(\varepsilon _{u}^{1,2})=1/2$ are
straightforwardly computed from the single subsystem field $F^{1}$ or $F^{2}$%
.\ On the other hand two outcome probabilities must take into account the
particle correlation (\ref{28}). It is thus impossible to obtain $%
\varepsilon _{u}^{1}=\varepsilon _{u}^{2}$: since there are no measurement
perturbations, $\varepsilon _{u}^{1}=\pm 1$ is associated with $\mathbf{r}%
_{1}\in \Sigma _{\pm u}^{1},$ implying $\mathbf{r}_{2}\in \Sigma _{\mp u}^{2}
$ so only $\varepsilon _{u}^{2}=\mp 1$ can be obtained. The probabilities in
this case read
\begin{equation}
P\left( \varepsilon _{u}^{1}=\pm 1,\varepsilon _{u}^{2}=\mp 1\right)
=P(\varepsilon _{u}^{1}=\pm 1)P(\varepsilon _{u}^{2}=\mp 1|\varepsilon
_{u}^{1}=\pm 1)=\frac{1}{2}  \label{100}
\end{equation}%
where the conditional probability is computed by way of the particle
dependence as $%\begin{equation}
%P_{\aleph}\left( \varepsilon_{a}^{1}=\pm1,\varepsilon_{a}^{2}=\pm1\right) =
P(\mathbf{r}_{1}\in \Sigma _{\pm u}^{1})P(\mathbf{r}_{2}\in \Sigma _{\mp
u}^{2}|\mathbf{r}_{1}\in \Sigma _{\pm u}^{1})$ and setting $b=u$ in Eqs. (%
\ref{41})-(\ref{42}). Note that these probabilities \emph{are not} equal to
those obtained by taking the relevant averages of $F_{T(u)}$ (eg, $%
\left\langle F_{T(u)}\right\rangle _{\Sigma _{+u}^{1}\Sigma _{+u}^{2}}$ does
not vanish).\ The reason is that $F_{T(u)}$ does not take into account the
particle correlation.\ It is possible nevertheless to identify the term
correlating the fields consistent with Eq. (\ref{28}) by rewriting Eq. (\ref%
{31}) as
\begin{equation}
F_{T(u)}(\mathbf{r}_{1},\mathbf{r}_{2})=F_{0(u)}(\mathbf{r}_{1},\mathbf{r}%
_{2})+e^{i\pi /2}F_{\aleph (u)}(\mathbf{r}_{1},\mathbf{r}_{2})  \label{56}
\end{equation}%
where $F_{0}$ and $F_{\aleph }$ are given by%
\begin{eqnarray}
F_{0(u)}(\mathbf{r}_{1},\mathbf{r}_{2})& =F_{\Sigma _{+u}}^{1}(\mathbf{r}%
_{1})F_{\Sigma _{+u}}^{2}(\mathbf{r}_{2})+F_{\Sigma _{-u}}^{1}(\mathbf{r}%
_{1})F_{\Sigma _{-u}}^{2}(\mathbf{r}_{2})  \label{58b} \\
F_{\aleph (u)}(\mathbf{r}_{1},\mathbf{r}_{2})& =F_{\Sigma _{+u}}^{1}(\mathbf{%
r}_{1})F_{\Sigma _{-u}}^{2}(\mathbf{r}_{2})-F_{\Sigma _{-u}}^{1}(\mathbf{r}%
_{1})F_{\Sigma _{+u}}^{2}(\mathbf{r}_{2}).  \label{58}
\end{eqnarray}%
It is easy to show that $F_{0(u)}$ cannot contribute to the probabilities by
repeating the reasoning involving no-perturbation measurements. On the other
hand $F_{\aleph (u)}$ respects by construction the particle correlation (\ref%
{28}) and the probabilities can be computed from the fields averages $%
\left\langle F_{\aleph }\right\rangle _{\Sigma _{\pm u}^{1}\Sigma _{\mp
u}^{2}}$\ (which are equal) and $\left\langle F_{\aleph }\right\rangle
_{\Sigma _{\pm u}^{1}\Sigma _{\pm u}^{2}}=0$. Note that the particle labels
as well as the field indices can be interchanged in the definition (\ref{31}%
) of $F_{T}$.

Let us now investigate measurements along arbitrary directions $a$ for
particle 1 and $b$ for particle 2.\ Probabilities for a single subsystem are
immediately obtained from the subsystem's field $F_{\alpha (u)\pm }^{1,2}$
yielding $1/2$ for any arbitrary direction. To compute correlations for two
outcomes, say $\varepsilon _{a}^{1}=1,\varepsilon _{b}^{2}=1$, the averages
involving $F_{T}$ must again be supplemented with the correlation (\ref{28}%
). This can be done by employing the equivalence relations $F_{\alpha (u)\pm
}\sim F_{\alpha (a)\pm }$ in Eq. (\ref{31}), yielding $F_{T(u)}\sim F_{T(a)}$%
. The probability is then computed as $P(\varepsilon
_{a}^{1}=1)P(\varepsilon _{b}^{2}=1|\varepsilon _{a}^{1}=1)$ by writing as
in Eq. (\ref{100}) single subsystem probabilities in terms of the particle
positions:%
\begin{eqnarray}
P_{T(a)}(\varepsilon _{a}^{1}=1,\varepsilon _{b}^{2}=1)& =P_{\alpha (a)+}(%
\mathbf{r}_{1}\in \Sigma _{+a}^{1})P_{\alpha (a)-}(\varepsilon _{b}^{2}=1|%
\mathbf{r}_{2}\in \Sigma _{-a}^{2})  \label{66} \\
& =\frac{1}{2}\sin ^{2}\left( \frac{\theta _{b}-\theta _{a}}{2}\right)
\label{66c}
\end{eqnarray}%
where we have used Eqs. (\ref{41}) and (\ref{42}). We can obviously reach
the same result by employing $F_{\alpha (u)\pm }\sim F_{\alpha (b)\pm }$ in
Eq. (\ref{31}) yielding%
\begin{equation}
P_{T(b)}(\varepsilon _{a}^{1}=1,\varepsilon _{b}^{2}=1)=P_{\alpha (b)-}(%
\mathbf{r}_{2}\in \Sigma _{+b}^{2})P_{\alpha (b)+}(\varepsilon _{a}^{1}=1|%
\mathbf{r}_{1}\in \Sigma _{-b}^{1}).  \label{68}
\end{equation}%
Both computations hinge on employing the form of the field that does not
perturb one of the measurements: this is necessary in order to be able to
compute conditional statements. As in the single particle system case [see
below Eq. (\ref{z1})] each particular realization of an equivalence class
gives rise to different, incompatible, accounts: Eq. (\ref{66}) specifies
that $\mathbf{r}_{1}\in \Sigma _{+a}^{1}$ while assuming the field
configuration is $F_{T(a)}$ whereas Eq. (\ref{68}) indicates that $\mathbf{r}%
_{1}\in \Sigma _{-b}^{1}$ when the field is $F_{T(b)}$. The direct
computation of $P_{T(u)}(\varepsilon _{a}^{1}=1,\varepsilon _{b}^{2}=1),$
without resorting to an equivalent configuration, cannot relie on
conditional statements since both measurements involve perturbations\footnote{%
Employing $P_{\alpha (u)+}(\mathbf{r}_{1}\in \Sigma _{+a}^{1})P_{\alpha
(u)-}(\varepsilon _{b}^{2}=1|\mathbf{r}_{2}\in \Sigma _{-a}^{2})$ along with
Eq. (\ref{42}) does not ensure the correlation is taken into account, since
one may have $\mathbf{r}_{1}\in \Sigma _{+a}^{1}$ and $\mathbf{r}_{2}\in
\Sigma _{-a}^{2}$ without $\mathbf{r}_{2}=-\mathbf{r}_{1}$.\ It is only when
at least one of the measurements is not perturbed that such an inference can
be made.}. The probability can be computed by obtaining the correlated
averages of the fields rotated by the interaction for each measurement. As
in the no-perturbation case $F_{\aleph (u)}$ is the field encapsulating the
correlation (\ref{28}) while $F_{0(u)}$ does not contribute to the
probabilities. This can be seen by noting that for the outcomes $\varepsilon
_{a,b}^{1,2}=1$ the averages $\left\langle F_{0(u),\aleph (u)}\right\rangle ,
$ giving $\cos (\frac{\theta _{b}-\theta _{a}}{2})$ and $\sin (\frac{\theta
_{b}-\theta _{a}}{2})$ for $F_{0(u)}$ and $F_{\aleph (u)}$ respectively do
not depend on $u$. This implies that $F_{0}$ and $F_{\aleph }$ form
separately \emph{equivalence classes}, i.e. we have%
\begin{equation}
F_{0(u)}\sim F_{0(a)}\mathrm{\hspace{.3cm}and\hspace{.3cm}}F_{\aleph (u)}\sim F_{\aleph (a)}
\label{62}
\end{equation}%
for any axes $u$ and $a$. Using Eq.\ (\ref{62}) it can be established that $%
F_{0}$ does not contribute to the probabilities\footnote{%
This can be seen by computing first $\left\langle F_{0}\right\rangle $ for
the outcomes $\varepsilon _{a}^{1,2}=1$ and $\varepsilon _{b}^{1,2}=1$ which
we know to vanish from the no-perturbation case (use $F_{0(a)}$ and $F_{0(b)}
$ respectively). The same averages can be computed instead from $F_{0(b)}$
and $F_{0(a)}$ implying that terms such as $\left\langle F_{\Sigma _{\lbrack
\pm b+a]}}^{1}\right\rangle _{+a}\left\langle F_{\Sigma _{\lbrack \pm
b+a]}}^{2}\right\rangle _{+a}$ must be put to zero by hand to take the
particle correlation into account.\ But these same terms also appear when
computing $P(\varepsilon _{a}^{1}=1,\varepsilon _{b}^{2}=1)$ from $%
\left\langle F_{0}\right\rangle $ with $F_{0(a)}$ and $F_{0(b)}$.}. Hence $%
P_{T(u)}$ is given by $P_{\aleph (u)}$%
\begin{equation}
 P_{\aleph (u)}(\varepsilon _{a}^{1}=1,\varepsilon _{b}^{2}=1)=\left\vert
\left\langle F_{\Sigma _{\lbrack u+a]}}^{1}\right\rangle _{+a}\left\langle
F_{\Sigma _{\lbrack -u+b]}}^{2}\right\rangle _{+b}-\left\langle F_{\Sigma
_{\lbrack -u+a]}}^{1}\right\rangle _{+a}\left\langle F_{\Sigma _{\lbrack
u+b]}}^{2}\right\rangle _{+b}\right\vert ^{2},
\end{equation}%
where the term between the $\left\vert ...\right\vert $ is the explicit
expression of $\left\langle F_{\aleph (u)}\right\rangle $. Of course, it is
possible (and simpler) to use Eq. (\ref{62}) and compute $P_{\aleph (a)}$ or
$P_{\aleph (b)}$ instead of $P_{\aleph (u)}$ (expressions similar to Eqs. (%
\ref{66}) and (\ref{68}) are obtained -- only the field indices need to be
changed despite $F_{\aleph }$ being defined jointly over the two subsystems).

\section{Discussion and conclusion}
The present dual field-particle model reproduces the EPR correlations
without the need to invoke non-locality (i.e. action at a distance): a
measurement carried out on one subsystem does not modify the field or the
particle position of the other system. A striking feature is that although
the total field $F_{T}$ is separable, the effective field $F_{\aleph }$ is a
non-separable function. Non-separability does not involve nor imply
non-locality (recall that non-separable functions are not exceptional in
classical physics\footnote{%
For example the classical action for a multi-particle system is a
non-separable function in configuration space.}) but is necessary in order
to account for field correlations between hemispheres encapsulating the
particle correlation (\ref{28}). The field configuration, as well as the
particle positions, are set at the source, in the intersection of the past
light cones of each subsystem's space-time location and are modified \emph{%
locally} by the measurement process\footnote{%
The non-separable part of the field $F_{T}$ that takes into account the
correlations between both subsystems becomes irrelevant to describe the
system once a measurement is made (since the correlations are broken at that
point).}. Several differences between our model and Bell-type LHV models
deserve to be pointed out. First, note that the probabilities are obtained
from average field intensities, not from elementary probabilities averaged
over HV distributions $\rho (\lambda )$.\ It is known that in general
classical fields do not have to obey Bell-type inequalities \cite{morgan}.
Here, the fields (i) are not necessarily positive valued, (ii) can
interfere, and (iii) define equivalence classes. Field measurements are
\emph{non-commutative}, whereas LHV models assume factorizable elementary
probabilities $p(\varepsilon _{a}^{1},\varepsilon _{b}^{2},\lambda
)=p(\varepsilon _{a}^{1},\lambda )p(\varepsilon _{b}^{2},\lambda )$, leading
to the existence of global joint probabilities (e.g. $P_{\aleph
}(\varepsilon _{a}^{1},\varepsilon _{a^{\prime }}^{1},\varepsilon
_{b}^{2},\varepsilon _{b^{\prime }}^{2})$ ) that in quantum mechanics can
only be defined for \emph{commuting} operators \cite{fine82}. If the
hemispherical fields were replaced by probability distributions for the
particles, then the equivalence relations would not hold and the conditional
probabilities appearing in Eqs. (\ref{66}) or (\ref{68}) would imply outcome
dependence \cite{matzkin jpa08,matzkin pra08}. The particles' positions thus
appear as pre-determined, and can play the role of hidden-variables, but
they do not ascribe probabilities except in the absence of measurement
perturbations. The field configurations can also be taken as hidden
variables and they do ascribe probabilities but only as members of an
equivalence class that does not give a more complete specification than
afforded by the quantum-mechanical state.

These last remarks lead us back to the original EPR dilemma recalled in the
Introduction. In a single particle system the field dynamics ensure that
there is no pre-existing outcome as an element of reality, even when it is
possible to make a prediction with unit probability (in this case too there
is an infinity of possible field-particle configurations, the outcome
arising from the interference between the system and the apparatus fields).
For an arbitrary measurement axis, a definite field-particle configuration,
even if known, would not give an elementary sub-quantum description of a
measurement outcome; such a description is only possible by resorting to an
equivalent, albeit fictitious, field particle configuration in which there
is no perturbation. In the two particle system, the additional constraint is
that the particle positions as well as the effective fields on each sphere
are correlated, allowing to infer one subsystem's outcome once the other
subsystem's outcome is known. This inference, in terms of a sub-quantum
description, also relies on the existence of an equivalence class providing
an equivalent configuration characterized by a no-perturbation measurement
along at least one axis. As a consequence the model denies the attribution
of simultaneous reality to $\varepsilon _{a}^{2}$ and $\varepsilon _{b}^{2}$
on the ground that an observer has the choice of measuring $\varepsilon
_{a}^{1}$ or $\varepsilon _{b}^{1}$ on particle 1 (this would imply that $%
F_{T(a)}$ and $F_{T(b)}$ be both realized as the system's field which is
impossible as noted above), although both conditional probabilities are
unity. Thereby the \textquotedblleft simultaneous reality\textquotedblright\
branch of EPR's dilemma -- which is fulfilled by Bell-type models -- is
decoupled here from the issue of locality.

To sum up, we have given an explicit model in which a quantum state appears
as an equivalence class comprising an infinity of possible field-particle
configurations.\ The model can be said to 'complete' quantum mechanics (in
the sense that it assumes an underlying reality relative to the quantum
state) though it does not generally allow to give more complete and
deterministic sub-quantum predictions. The model despite being local does
not abide by Bell's causality condition \cite{bell-loc} but nevertheless
defuses the EPR dilemma while avoiding the type of probability ascription
leading to Bell's theorem.

\end{document}